\documentclass[fleqn,12pt]{article}
\usepackage{epsfig}
\begin{document}                
\title{ Wigner distribution transformations in high-order systems }
\author{Jos\'e B. Almeida\\
\small{\emph{Universidade do Minho, Physics Department, 4710-057
Braga, Portugal}}\\ Vasudevan Lakshminarayanan\\
\small{\emph{University of Missouri - St. Louis, School of
Optometry}} \\  \small{\emph{and Department of Physics and
Astronomy, St. Louis, MO 63121, USA}}}
\date{}

%
\maketitle
\begin{abstract}                
By combining the definition of the Wigner distribution function
(WDF) and the matrix method of optical system modeling, we can
evaluate the transformation of the former in centered systems
with great complexity. The effect of stops and lens diameter are
also considered and are shown to be responsible for non-linear
clipping of the resulting WDF in the case of coherent
illumination and non-linear modulation of the WDF when the
illumination is incoherent. As an example, the study of a single
lens imaging systems illustrates the applicability of the method.
\end{abstract}
\section{Introduction}
 The Wigner distribution function (WDF) provides a
convenient way to describe an optical signal in space and spatial
frequency \cite{Bastiaans78, Dragoman97, Bastiaans97}. The
propagation of an optical signal through first-order optical
systems is well described by the WDF transformations
\cite{Bastiaans97, Bastiaans79, Bastiaans91}, allowing the
reconstruction of the propagated signal. Real optical systems are
not first order and the use of the WDF for optical system design
presumes the ability to predict how it is transformed by systems
with aberrations. Almeida \cite{Almeida99} proposed a method to
determine the aberration coefficients for optical systems using
matrix methods and calculated the necessary coefficients for
7th-order modeling of centered systems based on spherical
surfaces The extension of the matrix method to cylindrical
surfaces has also been proposed \cite{Laksh97}. Based on the fact
that the WDF lies between Fourier and geometric optics, we show
that geometric optics matrix coefficients can be used to predict
WDF transformations and hence can play an important role in
optical system design.
\section{Transformation of the WDF}
The Wigner distribution function (WDF) of a scalar, time
harmonic, and coherent field distribution $\varphi(\mathbf{q},z)$
can be defined at any arbitrary $z=\mathrm{const.}$ plane in terms
of either the field distribution or its Fourier transform
$\overline{\varphi}(\mathbf{p})=\int
\varphi(\mathbf{q})\exp(-ik\mathbf{q}^T \mathbf{p})d\mathbf{q}$
\cite{Bastiaans79, Dragoman97}:
\begin{eqnarray}
 \label{eq:Wigner}
 W(\mathbf{q},\mathbf{p})&=&\int
 \varphi\left(\mathbf{q}+\frac{\mathbf{q}'}{2}\right)
 \varphi^*\left(\mathbf{q}-\frac{\mathbf{q}'}{2}\right)
 \exp\left(-i k \mathbf{q}'^T \mathbf{p}\right)d \mathbf{q}' \\
 &=& \frac{k^2}{4\pi^2}\int
 \overline{\varphi}\left(\mathbf{p}+\frac{\mathbf{p}'}{2}\right)
 \overline{\varphi}^*\left(\mathbf{p}-\frac{\mathbf{p}'}{2}\right)
 \exp\left(i k \mathbf{q}^T \mathbf{p}'\right)d \mathbf{p}'~,
\end{eqnarray}
where $\mathbf{q}$ is the position vector, $\mathbf{p}$ the
vector of the conjugate momenta, $k=2 \pi/ \lambda$ and $^*$
indicates complex conjugate. In the present work we will be using
mainly \emph{quasi-homogeneous} light, in which case the WDF can
be defined as \cite{Bastiaans78, Bastiaans97}
\begin{equation}
    \label{eq:incoherent}
    W(\mathbf{q}, \mathbf{p}) = i(\mathbf{q}) \overline{s}(\mathbf{p}),
\end{equation}
where $i(\mathbf{q})$ is a non-negative function which we call
the intensity and $\overline{s}(\mathbf{p})$ is the Fourier
transform of the positional power spectrum $s (\mathbf{q})$ and
is also non-negative.

 If the position coordinates are
$x, y, z$ and the ray direction cosines are $u, v, w$, the
position and conjugate momenta vectors are given by
\begin{eqnarray}
    \label{eq:coords}
    \mathbf{q}&=& \left(\begin{array}{c}
      x \\
      y \
    \end{array}\right),\\
    \mathbf{p} &=& n \left(\begin{array}{c}
      u \\
      v \
    \end{array}\right),
\end{eqnarray}
where $n$ is the refractive index of the optical medium.

We will assume that the optical system is characterized by a
transfer map between the initial phase space coordinates,
$\mathbf{q}^i$, $\mathbf{p}^i$ and the final ones, $\mathbf{q}^f$,
$\mathbf{p}^f$. If $\mathcal{M}$ represents the transfer map:
\begin{equation}
 \label{eq:map}
 \left(\begin{array}{c}
   \mathbf{q}^f \\
   \mathbf{p}^f \
 \end{array}\right) = \mathcal{M}\left(\begin{array}{c}
   \mathbf{q}^i \\
   \mathbf{p}^i \
 \end{array}\right).
\end{equation}
We will also write expressions like
$\mathbf{q}^f=\mathcal{M}\mathbf{q}^i$ or $x^f = \mathcal{M} x^i$
to represent the dependencies of each of the final coordinates on
the original ones.

The transfer map can always be inverted; a simple physical
argument is sufficient to prove it: The transfer map is the
relationship between the ray coordinates on the input plane
($\mathbf{p}^i, \mathbf{q}^i$) and the corresponding coordinates
on the output plane ($\mathbf{p}^f, \mathbf{q}^f$); if two rays
share the same coordinates on the output plane they are the same
ray and so it is always possible to map the output onto the
input. This type of reasoning is valid within the scope of
geometrical optics, which corresponds to the conditions for
existence of a transfer map.

Having established that the map can be inverted the WDF
transformation is governed by the equation
\begin{equation}
 \label{eq:wigneralt}
 W^f(\mathbf{q}^f,\mathbf{p}^f) = |h(\mathbf{q}^f,\mathbf{p}^f)|
 W^i(\mathcal{M}^{-1}\mathbf{q}^f,\mathcal{M}^{-1}\mathbf{p}^f),
\end{equation}
where the factor $|h(\mathbf{q}^f,\mathbf{p}^f)|$ accounts for the
energy conservation between input and output and is the ratio
between the elementary hypervolume in input phase space and the
corresponding mapped hypervolume in output phase space:
\begin{equation}
    \label{eq:factor}
    |h(\mathbf{q}^f,\mathbf{p}^f)|=\left|\left(\frac{n^i}{n^f}\right)^2
    \frac{d x^i d y^i d u^i d v^i}{d x^f d y^f d u^f d
    v^f}\right|;
\end{equation}
if $\mathbf{J}$ is the jacobian of the map transformation we can
write \cite{Riley98}
\begin{equation}
 \label{eq:factor2}
 |h(\mathbf{q}^f,\mathbf{p}^f)|=\frac{1}{|\mathbf{J}|}.
\end{equation}

Eq.\ (\ref{eq:wigneralt}) is of special interest when the
transfer map can be expressed in closed form, which is the case if
matrices are used \cite{Kondo96, Laksh97, Almeida99}. In this
situation the output coordinates are expressed as polynomials in
the input coordinates or vice-versa. Almeida \cite{Almeida99}
showed that this method can be extended to any desired degree of
approximation, at least for centered systems, and published the
coefficients for the 7th-order matrices of systems based on
spherical surfaces.

There are two methods of map inversion in matrix optics. The first
one is a straightforward matrix inversion and can be used in many
cases; the second one, applicable in all circumstances, consists
in reversing the optical system and recalculating all the matrix
coefficients. We can thus find $\mathcal{M}$, $\mathcal{M}^{-1}$
and $\mathbf{J}$ for any centered optical system, no matter how
complex. There remains a question about the aperture stops which
is dealt with below.

In order to model a system with matrices we start by defining a
generalized ray of complex coordinates $Q^i = x^i+ jy^i$ and $P^i
= n(u^i +j v^i)$; this ray is described by the 40-element
monomials vector $\mathbf{Q}^i\&$, built according to the rules
explained by Kondo et al. \cite{Kondo96} and Almeida
\cite{Almeida99}. If the ray is subjected to a transformation
described by matrix $\mathbf{M}$, then the output ray has
coordinates $(Q^f, P^f)$ and is represented by the monomials
vector $\mathbf{Q}^f\&$, such that:
\begin{equation}
  \label{eq:matrixmap}
  \mathbf{Q}^f\& = \mathbf{M Q}^i\&.
\end{equation}

For an axis symmetric optical system, in the 7th-order, matrix
$\mathbf{M}$ will result from a product of $40\times40$ square
matrices with real elements. Each matrix in the product describes
a specific ray transformation. The elementary transformations can
be classified into four different categories:

\begin{description}
  \item {\textbf{Translation:}} A straight ray path.
  \item {\textbf{Surface refraction:}} Change in ray orientation governed by Snell's law.
  \item {\textbf{Forward offset:}} Ray path between the surface vertex plane and the
  surface.
  \item {\textbf{Reverse offset:}} Ray path from the surface back to the vertex plane,
  along the refracted ray direction.
\end{description}

The ray itself is described by a 40-element vector comprising the
monomials of the complex position and conjugate momenta
coordinates that have non-zero coefficients; the first two
elements of this vector are just the complex coordinates $(Q,P)$.

Considering Eq.\ (\ref{eq:matrixmap}) the mapping equation
(\ref{eq:map}) takes the form
\begin{eqnarray}
    \label{eq:pols1}
    Q^f &=& \mathcal{P}_Q (Q^i, Q^{i*}, P^i, P^{i*});  \\
    \label{eq:pols2}
    P^f &=& \mathcal{P}_P (Q^i, Q^{i*}, P^i, P^{i*});  \\
    \label{eq:pols3}
    Q^i &=& \mathcal{P}_Q^{-1} (Q^f, Q^{f*}, P^f, P^{f*});  \\
    \label{eq:pols4}
    P^i &=& \mathcal{P}_P^{-1} (Q^f, Q^{f*}, P^f, P^{f*}).
\end{eqnarray}
The symbols $\mathcal{P}_Q$, $\mathcal{P}_P$,
$\mathcal{P}_Q^{-1}$, $\mathcal{P}_P^{-1}$ mean polynomial
expressions of the variables in parenthesis. Eqs.\
(\ref{eq:pols3}) and (\ref{eq:pols4}) can now be used to evaluate
Eq.\ (\ref{eq:factor2}) first and then Eq.\ (\ref{eq:wigneralt}).
\section{Stops and pupils}
Any system analysis is incomplete without consideration of the
effect of the various stops along the optical path; this analysis
cannot be incorporated in the matrix description and deserves
special treatment. Paraxial theory tells us \cite{Born80} that we
can find one most limiting stop whose images in object and image
space are known by entrance and exit pupils, respectively. The
theory goes that the entrance pupil establishes the width of the
beam entering the system while the field angle is established by
the second most limiting stop imaged onto object space; the
images of the same stops in image space set corresponding limits
to the rays leaving the system. It is not necessary to leave
paraxial theory to find that these concepts are insufficient for
the complete description of the beam constraints within the system
and we are led to the concept of vignetting.

Moving from the paraxial approximation to high-order the problem
increases in complexity and even the concepts of entrance and exit
pupil loose significance in view of the high aberrations present
when an internal stop is imaged to either object or image space
\cite{Almeida99:3}. Ray tracing software usually avoids the
problem by imposing restrictions as the rays cross each stop's
plane \cite{Oslo95}.

In order to tackle the problem in phase space, we will define
\emph{scene} as an optical field distribution that spans $-\infty
< |\mathbf{q}| < +\infty$ in space coordinates and $-n <
|\mathbf{p}| < n$ in conjugate momenta. A scene cannot contain
components with $|\mathbf{p}| \geq n$ because these components
originate evanescent waves that are considered faded out
\cite{Goodman68, Almeida00:3}. What the paraxial theory says is
that the entrance pupil clips the scene in $\mathbf{p}$
coordinates, while the second most limiting stop is responsible
for clipping in $\mathbf{q}$ coordinates. If only the meridional
plane is considered, to reduce the dimensions to 2, the stops
create a parallelogram area in phase space, with two sides
parallel to the $p$ axis, where the WDF is non-zero. Vignetting
must be seen as a departure from that form, meaning that, for the
extreme values of $\mathbf{q}$, the angular spread of the rays may
be different from the central one.

In order to understand the stop effects on the WDF we consider it
as a modulator, in which case the following relation applies
\cite{Bastiaans96}:
\begin{equation}
    \label{eq:modulator}
    W^f (\mathbf{q}, \mathbf{p}) = \frac{k^2}{4 \pi^2}
    \int W^m(\mathbf{q}, \mathbf{p}-\mathbf{p}') W^i(\mathbf{q},
    \mathbf{p}')d \mathbf{p}',
\end{equation}
where $W^m(\mathbf{q}, \mathbf{p})$ is the WDF of the modulating
function $m(\mathbf{q})$. Eq.\ (\ref{eq:modulator}) represents a
two-dimensional convolution of the Wigner distribution functions
$W^m(\mathbf{q}, \mathbf{p})$ and $W^i(\mathbf{q}, \mathbf{p})$
with respect to the frequency variables and a  mere
multiplication with respect to the space variables.

A stop is a special kind of modulator. In coherent illumination
the stop has a modulating function that equals unity within the
stop area and is zero elsewhere. Furthermore, as we are usually
dealing with stops that are very large compared to the
wavelength, Eq.\ (\ref{eq:modulator}) results in clipping of the
local WDF in the space domain:
\begin{eqnarray}
    \label{eq:stop}
    W^f (\mathbf{q}, \mathbf{p})=W^i (\mathbf{q}, \mathbf{p}),&&
    \mathrm{within~the~stop}, \nonumber \\
    W^f (\mathbf{q}, \mathbf{p})=0, && \mathrm{elsewhere}.
\end{eqnarray}
In incoherent illumination the stop modulating function is the
auto-correlation function of the stop transmittance function
\cite{Born80, Goodman68}. If as before the stop dimensions are
large compared to the wavelength, Eq.\ (\ref{eq:modulator}) can
be written
\begin{eqnarray}
    \label{eq:stopinc}
    W^f (\mathbf{q}, \mathbf{p}) = W^i (\mathbf{q}, \mathbf{p})
    S(\mathbf{q}),&&
    \mathrm{within~the~stop}, \nonumber \\
    W^f (\mathbf{q}, \mathbf{p})=0, && \mathrm{elsewhere}.
\end{eqnarray}
The stop auto-correlation function is defined by
\begin{equation}
    \label{eq:sacf}
    S(\mathbf{q})= \int S(\mathbf{q}+\mathbf{q}')S(\mathbf{q})d
    \mathbf{q}'.
\end{equation}

The translation of the stop modulation onto an equivalent effect
of the original scene's WDF depends on the sort of transformations
the latter has incurred up to that point. When the signal
encounters the first stop in the system the only transformation
that the WDF has suffered is a spatial shearing, which is linear
in the paraxial approximation and non-linear if wide angles are
considered \cite{Almeida00:3}. If the distance from the scene to
the stop is large the angle subtended by the stop will be
virtually independent from the position coordinates on the scene
and the stop effect will be virtually equivalent to a clipping or
modulation on the spatial frequency domain. This is what an
entrance pupil is supposed to do and so we state that an entrance
pupil is a concept valid in the paraxial approximation, when the
scene is very far from the optical system.

The effect of further stops along the system is more difficult to
understand. Let us assume that we are dealing with small angles,
such that paraxial approximation is indeed applicable, that we
have converted the existing stops to their equivalents in object
space and let us consider just the two most significant ones. The
problem has been reduced to free-space propagation, characterized
by linear shearing of the WDF.

\begin{figure}[p]
\caption{The effect of stops on propagation angles. S$_2$ is the
entrance pupil and object point P$_2$ is on the edge of the
field.}
    \vspace{20mm}
    \centerline{\psfig{file=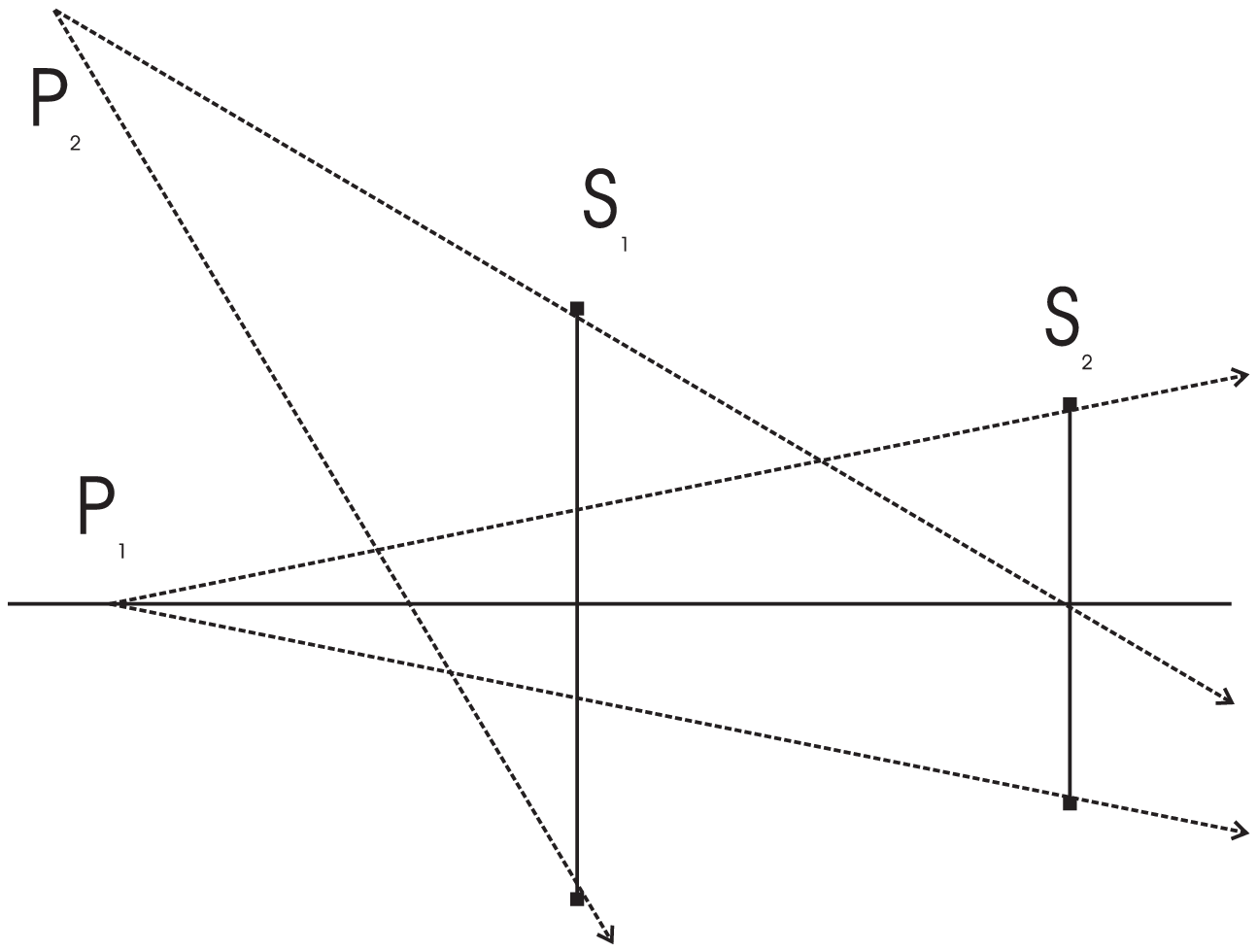, scale=0.8}}
\label{fig:stops}
\end{figure}

Fig.\ \ref{fig:stops} illustrates the situation described above;
object point P$_1$ is an axis point and obviously stop S$_2$ is
the entrance pupil, responsible for the limitation on the rays
that enter the system. According to general practice, we would
define the field limits as the points on the rays that pass on
the edges of the stop S$_1$ and the center of the entrance pupil;
point P$_2$ is one such point. Naturally, stop S$_1$ also
introduces limitations on the rays that enter the system, besides
its prime function as field limiter; this effect is known as
vignetting.

\begin{figure}[p]
\caption{Effect of stops in phase space. a) Clipping at the
entrance pupil, b) the WDF is back-propagated from the entrance
pupil to the field pupil and clipped, c) back-propagation to the
scene plane.}
    \vspace{10mm}
    \mbox{a) \psfig{file=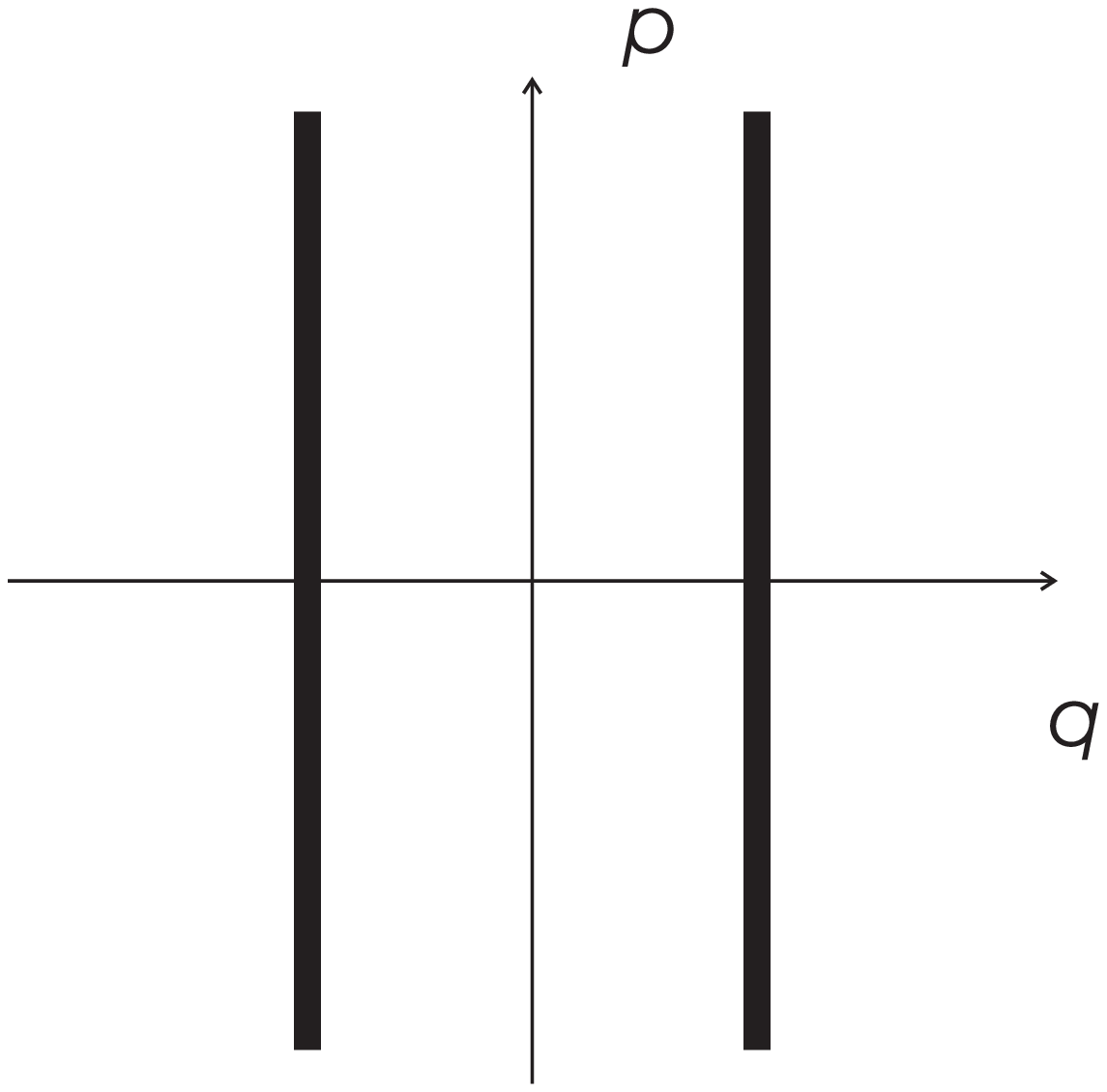, scale=0.4}}
    \mbox{b) \psfig{file=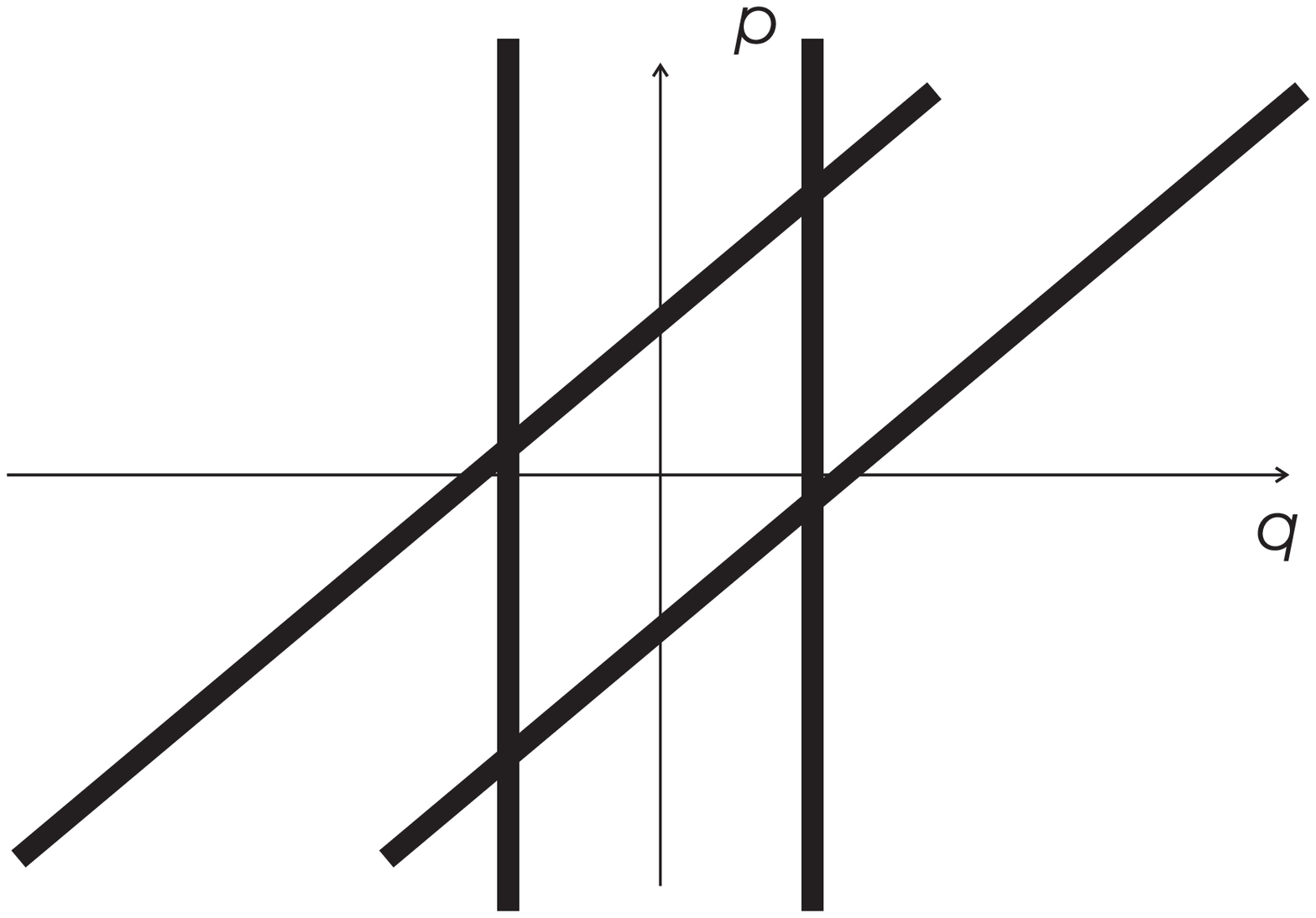, scale=0.4}}\\ \\ \\
    \mbox{c) \psfig{file=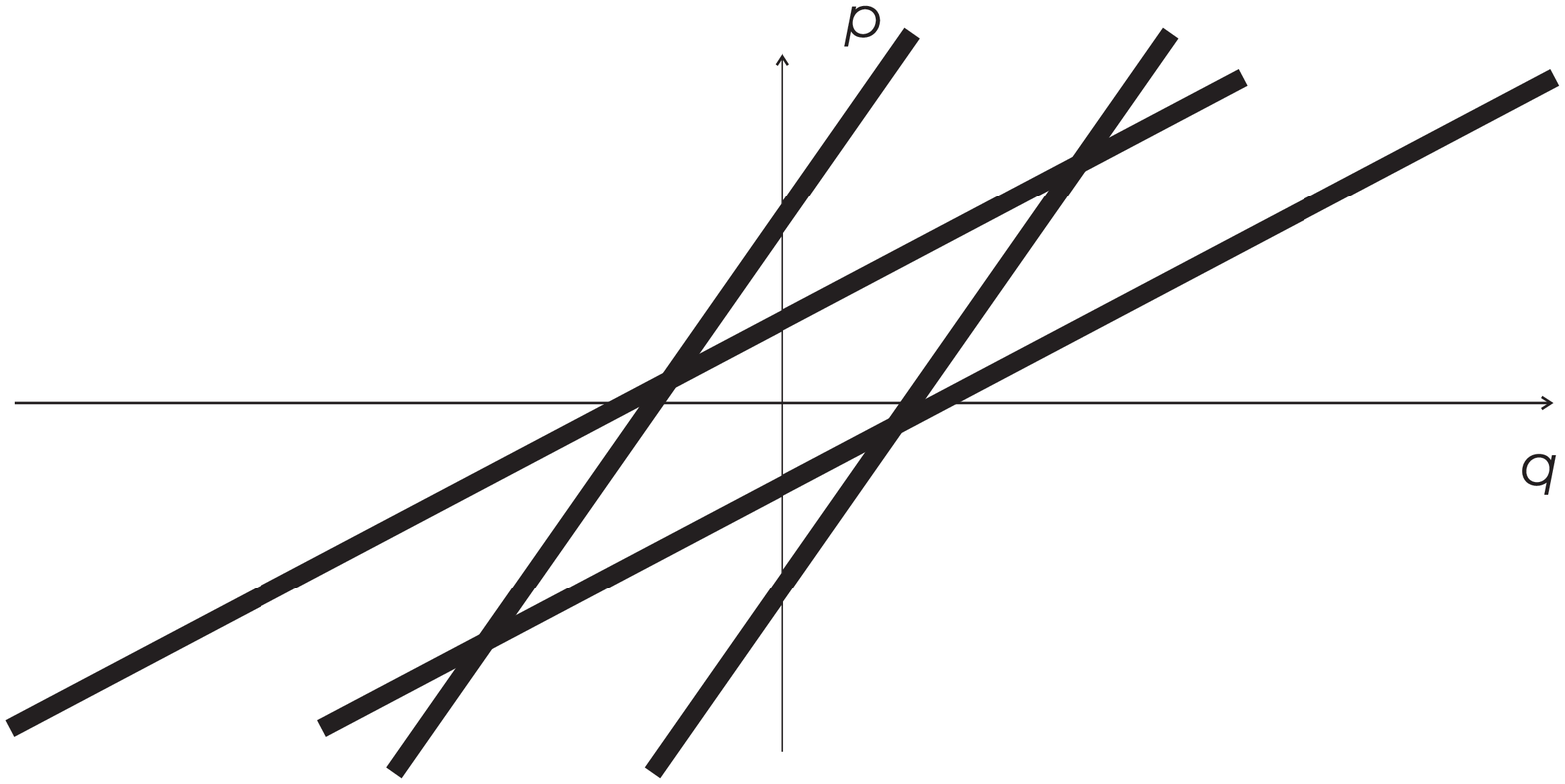, scale=0.4}}
\label{fig:phasestop}
\end{figure}

The effect in phase space is illustrated in Fig.\
\ref{fig:phasestop} where a similar situation is depicted. Fig.\
{\ref{fig:phasestop} a)} shows the effect of the entrance pupil on
the local WDF with a clipping to its own width. The field pupil,
stop S$_1$, produces its clipping on the WDF back-propagated from
the entrance pupil, as shown in Fig.\ {\ref{fig:phasestop} b)},
and the resulting double-clipped WDF is back-propagated to the
scene plane, as shown in Fig.\ {\ref{fig:phasestop} c)}. The
resulting parallelogram shape is the representation in phase
space of the signal that can, in fact, enter the system; it is
clear that for a point on the axis, $q=0$, the stop S$_2$ is
responsible for determining the admittance angle, while stop
S$_1$ is, to a great extent, responsible for determining the
dominion of $q$, which is exactly what we call field. The
vignetting effect is visible for extreme values of $q$, for which
stop S$_2$ no longer determines completely the admittance angle.

The extension of the above procedures to a general mapping
situation, outside the paraxial approximation, must be done
carefully. In the next section we study a complete mapping
situation, illustrating both the WDF transformations and stop
consideration.

\section{Example}
The case below was chosen not for its particular applicability
but for its ability to demonstrate and highlight the
possibilities opened by the matrix mapping and WDF used together.

We will consider a simple imaging system composed of a single
convex lens and a field stop on the image plane. The lens was
chosen to produce a high degree of aberrations, so that the
non-linear effects are clearly visible. The lens is plano-convex,
with the flat surface facing the image plane, and has a
refractive index of $1.56$; the convex surface has a radius of
$6.5 \times 10^{-2}~\mathrm{m}$ and the central thickness is
$2.0\times 10^{-2}~\mathrm{m}$. The lens and field stop diameters
will be decided later on, upon examination of the aberrations
present in the image.

Eqs.\ (\ref{eq:pols1} to \ref{eq:pols4}) were established for the
system in consideration using the method outlined by Almeida
\cite{Almeida99} using Mathematica \cite{mathematica}. The same
software package was also used for all the further calculations.
The analysis was carried out on a meridional plane, so the
detection of aberration effects such as astigmatism is out of the
question. The input scene was defined according to Eq.\
(\ref{eq:incoherent}) in terms of its WDF as:
\begin{equation}
    \label{eq:inputscene}
    W^i(q,p)= 1 + \sin \frac{4 \pi q}{l},
\end{equation}
where $l$ is a parameter used to control the detail on the scene;
all the graphics were plotted with $l=5 \times
10^{-2}~\mathrm{m}$.

\begin{figure}[p]
\caption{Transformation of the Wigner distribution function
through a lens; a) input distribution, b) output distribution.
Both figures show the clipping effect of the lens diameter
(dashed line) and the field stop (solid line).}
    \vspace{10mm}
    \mbox{a) \psfig{file=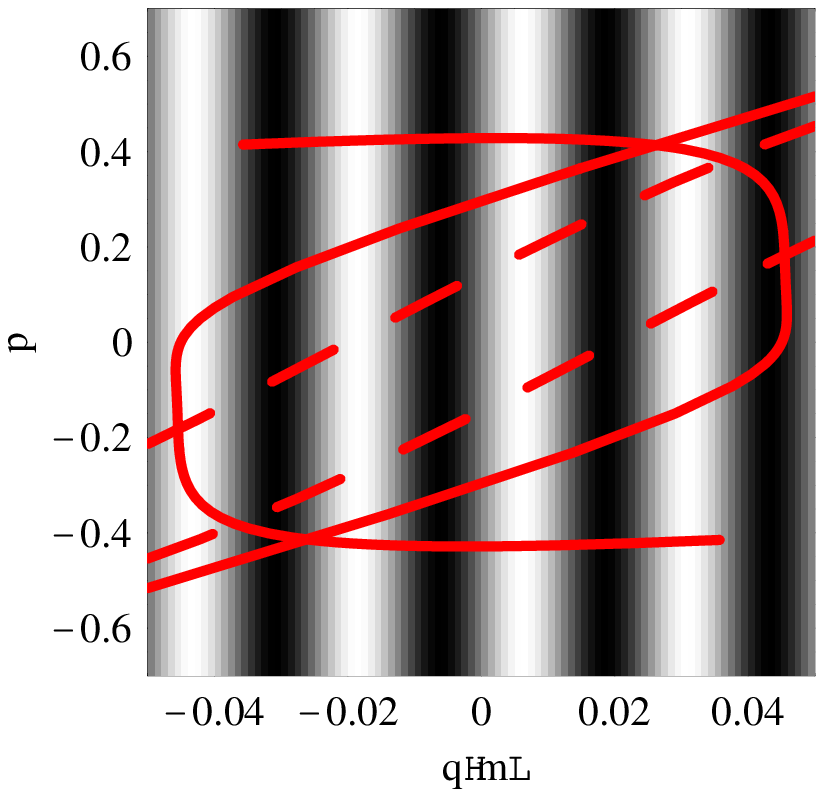, scale=0.7}}
    \mbox{b) \psfig{file=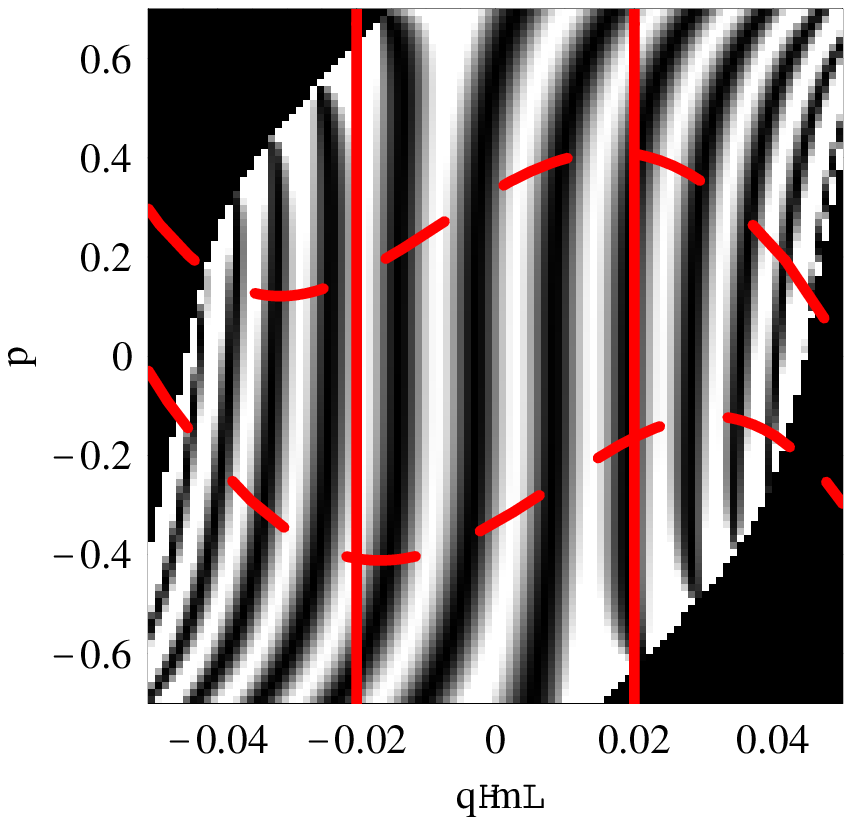, scale=0.67}}
\label{fig:wignerpup}
\end{figure}

The input scene was located at $13.7 \times 10^{-2}~ \mathrm{m}$,
so that the image was formed at $5 \times 10^{-2}~ \mathrm{m}$.
Fig.\ \ref{fig:wignerpup} shows the input scene in phase space
and the output WDF. The input appears as a series of light and
dark bands, showing the independence of the corresponding WDF on
the $p$ coordinate, characteristic of \emph{spatially incoherent}
light, a special case of \emph{quasi-homogeneous} light
\cite{Bastiaans78, Bastiaans97}. The output shows the same bands,
reduced in width due to a magnification factor lower than unity
and distorted by aberrations. A qualitative analysis of the
aberrations is indeed interesting.

The S-shape of the bands results from spherical aberration of
various orders, with predominance of the third-order. The reduced
width of the bands for higher values of $q$ is characteristic of
barrel distortion; this so high that the signal does not exist
above $|q|> 0.5$, except for the effect of spherical aberration.
This is similar to a fish-eye objective. Field curvature is
clearly visible as tilting of the central portion of the bands
for high $|q|$. Coma results in an asymmetry of the S-shape.

Clearly we have performed a mapping with an infinite diameter
lens, which only works mathematically. Considering the radius and
width of the curved surface, we have established a lens diameter
of $4 \times 10^{-2}~ \mathrm{m}$. The lens diameter was given to
a stop located on the vertex plane and the edges of this stop were
mapped forward, through the lens and free-space, to the image
plane, and backward to the input scene plane. The maps of the
lens diameter stop are superimposed on the corresponding WDFs as
dashed lines. We decided to use a field stop on the image plane,
in order to limit the image to an area of low aberration; a
diameter of $4 \times 10^{-2}~ \mathrm{m}$ was also chosen for
this stop. The field stop was mapped onto the input scene plane
and is shown as a solid line superimposed on both figures. If we
were dealing with coherent light the area of both WDFs common to
the zones defined by the two stops would be the area relevant for
the image formation; in fact, to put it correctly, the image WDF
should have been made equal to zero outside that area.

\begin{figure}[p]
\caption{The Wigner distribution function on the image plane
shows the effect of stops. With incoherent illumination the stops
don't clip the distribution but apply a smoothing from the center
of the stop to twice the stop width. }
    \vspace{20mm}
    \centerline{\psfig{file=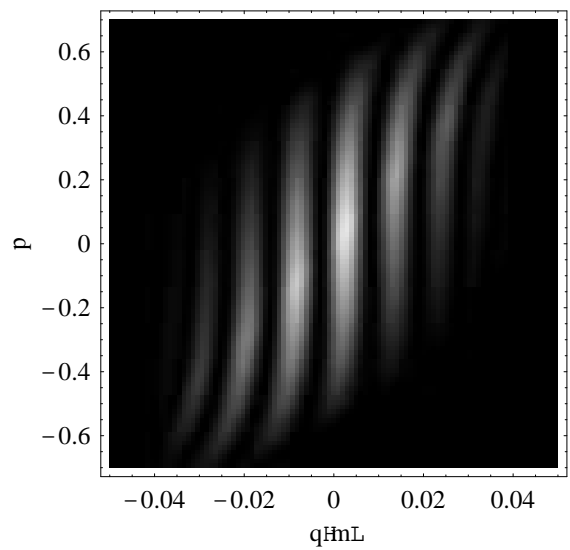, scale=0.8}}
\label{fig:wignersmooth}
\end{figure}

For a one-dimensional stop Eq.\ (\ref{eq:sacf}) becomes
\cite{Goodman68}
\begin{eqnarray}
    \label{eq:sacf1D}
    S(q) = 1 - \frac{|q|}{ 2d}, &\rightarrow& |q|\leq 2 d,
    \nonumber \\
    S(q)=0, &\rightarrow& |q|> 2 d.,
\end{eqnarray}
where $d$ is the half-width of the stop. So, in incoherent
illumination, rather than clipping the local WDF, the stop
produces a gradual transition from full intensity to zero with
twice the width of the stop. When propagated to either the image
or the object planes this transition manifests itself as a gradual
transition of the WDF from the full mapped value to zero guided
but not delimited by the stops' traces, see Fig.{\ }
\ref{fig:wignersmooth}.
\section{System analysis}
Although not presented in this paper, it would be possible to
extract a lot of information about the system from the image WDF
modulated by the stops. The field distribution would be obtainable
directly from an integration of the image WDF in the variable
$p$; the integration limits would be established by maps of stops
twice the width of real the stops. It is clear that, within the
region delimited by the field stop, there is a reasonable
reproduction of the original scene.

The point spread function for an input point $q_0$ could be
evaluated considering a different input scene, such as $W^i
(q,p)= \delta (q-q_0)$, and again integrating the output WDF in
$p$. The MTF could also be evaluated using the same scene but
performing the integration in $q$.
\section{Conclusions}
The authors presented a method to evaluate the WDF transformations
of an optical signal that passes through a system, in the context
of geometrical optics. Using matrices it is possible to model
centered systems up to any desired order of approximation; the
authors have shown that the same matrix method can be used for
the evaluation of the WDF transformations.

The effect of stops and lens diameter could also be accounted for
leading to the definition of clipping traces on both the input
and output WDF and to the outline of methods to evaluate the
resulting field distribution, point spread function and MTF.
\section{Acknowledgements}
J. B. Almeida wishes to acknowledge the fruitful discussions with
P. Andr\'es, W. Furlan and G. Saavedra at the University of
Valencia.

On the occasion of his retirement, this paper is dedicated to
Professor K Srinivasa Rao in celebration of his long productive
career  in theoretical physics.
 \pagebreak
  \bibliography{aberrations}   
  \bibliographystyle{OSA}   

\end{document}